\documentclass[a4paper,11pt]{article}
\usepackage{pos}

\title{New results from the DANSS experiment}

\author*[a]{Mikhail Danilov}

\affiliation[a]{Lebedev Physical Institute of the Russian Academy of Sciences,\\
  53 Leninskiy Prospekt, Moscow, Russia
 \\On behalf of the DANSS Collaboration}

\emailAdd{danilov@lebedev.ru}

\abstract{
There are several experimental indications of sterile neutrinos with a mass in the 1~eV ballpark and many experiments are trying to clarify the situation. 
During 6 years the DANSS experiment collected more than 6 million Inverse Beta Decay (IBD) events and measured the background level during 4 reactor-off periods. Data were collected at 3 distances (10.9~m, 11.9~m, and 12.9~m) from the  center of the core of a $3.1$~GW$_{th}$ reactor with event rate up to 5 thousand per day. The detector position was changed frequently usually 2-3 times a week. Therefore many systematic uncertainties were canceled out in the analysis. After collection of additional 0.7 million IBD events  
the  significance of the best-fit point in the 4$\nu$ case increased from  1.3$\sigma$ to 2.35$\sigma$. This is still not statistically significant and we present 
the exclusion area that covers a very interesting range of the sterile neutrino parameters up to $\sin^22\theta_{ee} < 0.004$ in the most sensitive point. In particular DANSS excludes a large fraction of sterile neutrino parameters preferred by the recent BEST results including the BEST best-fit point. 
The IBD rate dependence on the fuel composition was measured. It agrees with predictions of the Huber-Mueller model. During almost 6 years the reactor power was measured  with $\approx1.8\%$ accuracy in 2 days  using the anti-neutrino event rate normalized to the reactor power at the initial period.
}

\FullConference{%
 The International Conference on High Energy Physics (ICHEP 2022)\\
  6-13 July 2022\\
  Bologna, Italy\\
}

\begin{document}
\maketitle

\def\mydm{\Delta m^2_{41}}
\def\mysin{\sin^2 2\theta_{ee}}
\def\oscillationspars{$\mydm$, $\mysin$}
\def\antiparticle{\tilde}
\def\antinu{$\tilde{\nu_e}$}
\def\clsmethod{CL$_s$}
\def\cl{\mathrm{CL}}

\section{Introduction}

The deficit of $\nu_e$ in the calibration of the SAGE and GALEX experimets with radioactive sources~\cite{SAGE, GALEX} (``Galium Anomaly''(GA)) and the deficit in reactor $\antiparticle\nu_e$ fluxes~\cite{Huber,Mueller} (``Reactor Antineutrino Anomaly''(RAA)) can be explained by active-sterile neutrino oscillations.

RAA can be explained by
new measurements~\cite{Kopeikin:2021rnb,Kopeikin:2021ugh} of beta spectra of fission products of $^{235}$U and $^{239}$Pu that give 5.4\%  smaller ratio than the ILL results used for predictions of reactor $\antiparticle\nu_e$~\ fluxes~\cite{Huber,Mueller}. 

Recently the BEST experiment confirmed GA~\cite{BEST,BEST-2022-arxiv} with more than 
5$\sigma$ significance. For $\Delta m^2_{41} < 5$eV$^2$ very large values of $\sin^2 2\theta_{ee}\approx 0.4$ preferred by BEST are excluded by DANSS~\cite{Alekseev:2018efk,Danilov:2020ucs} and NEOS~\cite{Ko:2016owz}. For $\Delta m^2_{41} > 5$eV$^2$ the BEST results are in tension
with several other experiments~\cite{BEST-2022-arxiv}.

The MiniBooNE collaboration 
confirmed the LSND results on 
$\nu_e$($\antiparticle\nu_e$) appearance in the 
$\nu_{\mu}$($\antiparticle\nu_{\mu}$)
beams with a combined significance of 6$\sigma$ ~\cite{MiniBooNE2}. 
Appearance of electron neutrinos in muon neutrino beams can be explained by  sterile neutrinos. The MicroBooNE collaboration has not confirmed the MiniBooNE results but has not excluded them completely~\cite{MicroBooNE:2021rmx}.

The Neutrino-4 experiment claimed an observation of  sterile neutrinos 
although the significance of the result was only 2.7$\sigma$~\cite{Serebrov:2018vdw,Serebrov:2020kmd} and there were concerns about the validity of their analysis (see e.g.~\cite{Danilov:2020rax}). 

The survival probability of reactor $\antiparticle\nu_e$ at very short distances  in the 4$\nu$ mixing scenario (3 active and 1 sterile neutrino) is given by the formula:\\
$$1-\mysin \sin^2({1.27\mydm [\mathrm{eV}^2] L[\mathrm m]}/{E_\nu [\mathrm{MeV}]}),$$
where $\mysin$ is the mixing parameter, $\mydm = m_4^2 - m_1^2$ is the difference in the squared masses of neutrino
mass states, $L$ is the distance between production and detection points and $E_{\nu}$ is the $\antiparticle\nu_e$ energy.
 The Inverse Beta Decay (IBD) reaction 
$\antiparticle{\nu}_e + p \rightarrow e^+ + n$ 
is used to detect $\antiparticle\nu_e$. In this reaction $E_{\nu}\approx E_{e^+} + 1.8$~MeV.
\section{Detector, data analysis and positron spectra.}
The DANSS detector~\cite{DANSS} is located on a movable platform under the core of  
a 3.1~GW$_{th}$ reactor of the Kalininskaya NPP in Russia.
 It is made of 2.5 thousand $100\times4\times1$~cm$^3$ scintillator strips with Gd-loaded surface coating. Strips in neighbor layers are orthogonal. This allows a quasi-3D reconstruction of events.  
Three wavelength-shifting fibers placed in grooves along the strip are used to collect the scintillation light. The central
fiber is read out with a SiPM and the two side fibers from 50 parallel 
strips are readout with a compact PMT. External backgrounds are suppressed by a multi-layer active and passive shielding. 
 The kinetic energy  of the reconstructed positron is used in the analysis that is 1.02~MeV smaller than the prompt energy used in other experiments.  This approach does not depend on energies of annihilation gammas that have a nonlinear energy response.

The SiPM's gains and cross-talks vary with temperature. They are calibrated every 25--30 minutes using noise signals collected in parallel with data taking. The energy responses of all 2500 scintillator counters are calibrated every 2 days using cosmic muons.  A median of the Landau distribution is used in the analysis since it is more stable than the most probable value.  

The detector energy scale is fixed with the  $^{12}$B beta spectrum 
measured using two $^{12}$B production mechanisms: $\mu^-$ captures on carbon 
and neutron interactions with carbon. 
The lifetime of $^{12}$B coincides nicely  with the expected $\tau=29.1$~ms in both cases ($28.6\pm0.7$~ms and $29.4\pm0.6$~ms correspondingly). The energy scales agree in both cases within $\pm0.2\%$.
The energy scale that is determined using  
$^{60}$Co and $^{248}$Cm radioactive sources agrees within $\pm0.2\%$ with the $^{12}$B scale. 
The response to the $^{22}$Na source is 1.8\% lower than expected probably because of many soft gammas from this source that are not well enough described by the DANSS Monte Carlo.
Such soft gammas are rare in positron and $^{12}$B events.
Nevertheless, we still use in the analysis a 2\% uncertainty in the energy scale.

 The accidental coincidence background 
is calculated in a model independent way using  time intervals for neutron signals preceding  the positron signal. 
Several cuts~\cite{DANSSdata,Svirida:2020zpk}
 reduce the accidental background to 15.3\% of the IBD signal at the top detector position for the positrons in the (1.5-6)~MeV energy interval used in the oscillation analysis. 
Other background estimates below use the same conditions.
The background from neighbor reactors is 0.6\% only and has a known shape.
Other correlated backgrounds were estimated using 4 reactor-off periods to be less than 1.8\%.

\begin{figure}[h]
\begin{center}
\includegraphics[width=0.98\textwidth]{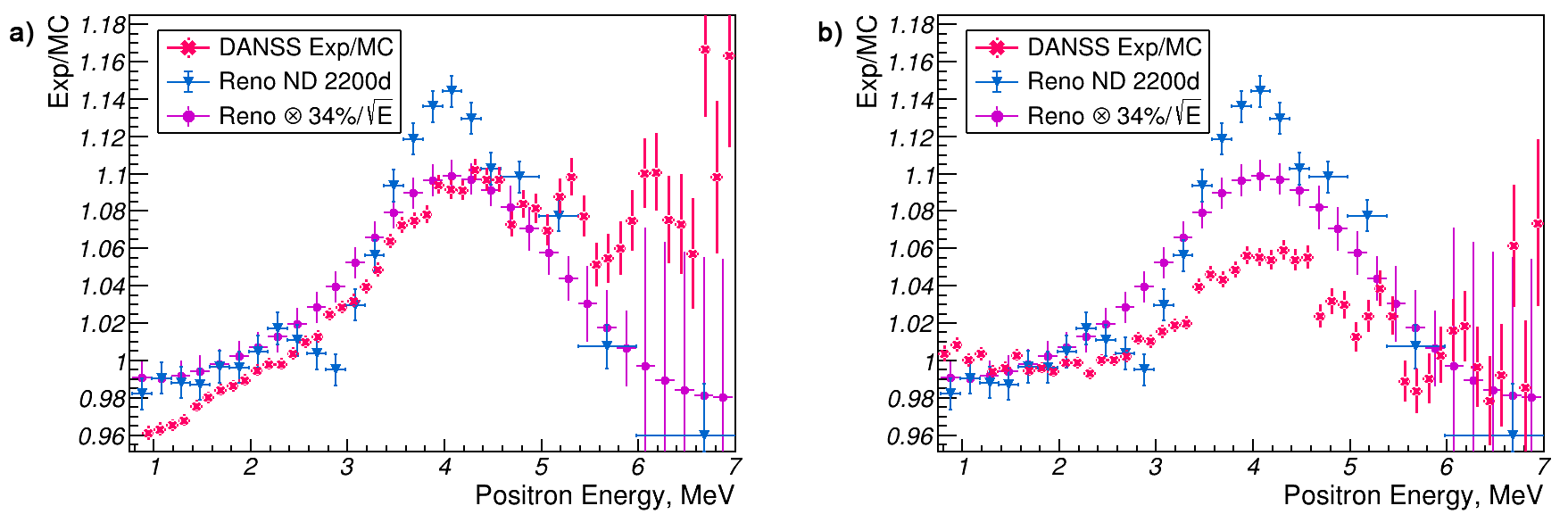}
\end{center}
\caption{\label{fig:bump} a) Experiment to the H-M model MC ratio for $e^+$ spectrum; b) the same ratio with 
$e^+$ energy  shifted by -50~keV.} 

\end{figure}

A ratio of the measured positron spectrum and the Huber-Mueller (H-M) model~\cite{Huber,Mueller} MC predictions  is shown in Fig.~\ref{fig:bump}a as well as the 
RENO results
~\cite{Bak:2018ydk} shifted by -1.02~MeV to take into account the difference between the $e^+$ kinetic energy used by DANSS and the prompt energy used by RENO. 
The RENO spectrum smeared by the DANSS
energy resolution is also shown. The best agreement between our measurements and the MC predictions
in the range $1.5 - 3.0$~MeV is obtained if the $e^+$ energy is shifted by -50~keV which is within the systematic uncertainties 
(Fig.~\ref{fig:bump}b). In this case we observe a bump similar in shape and position with the RENO bump smeared by the DANSS resolution. However the height of the bump is considerably smaller.  
Because of the high sensitivity of the measured ratio to the energy scale and shifts we can not claim the existence  of the bump. The results of other experiments which claim the existence of the bump should be also very sensitive to the energy scale and possible shifts.  

\begin{figure}[h]
\begin{center}
\includegraphics[width=0.98\textwidth]{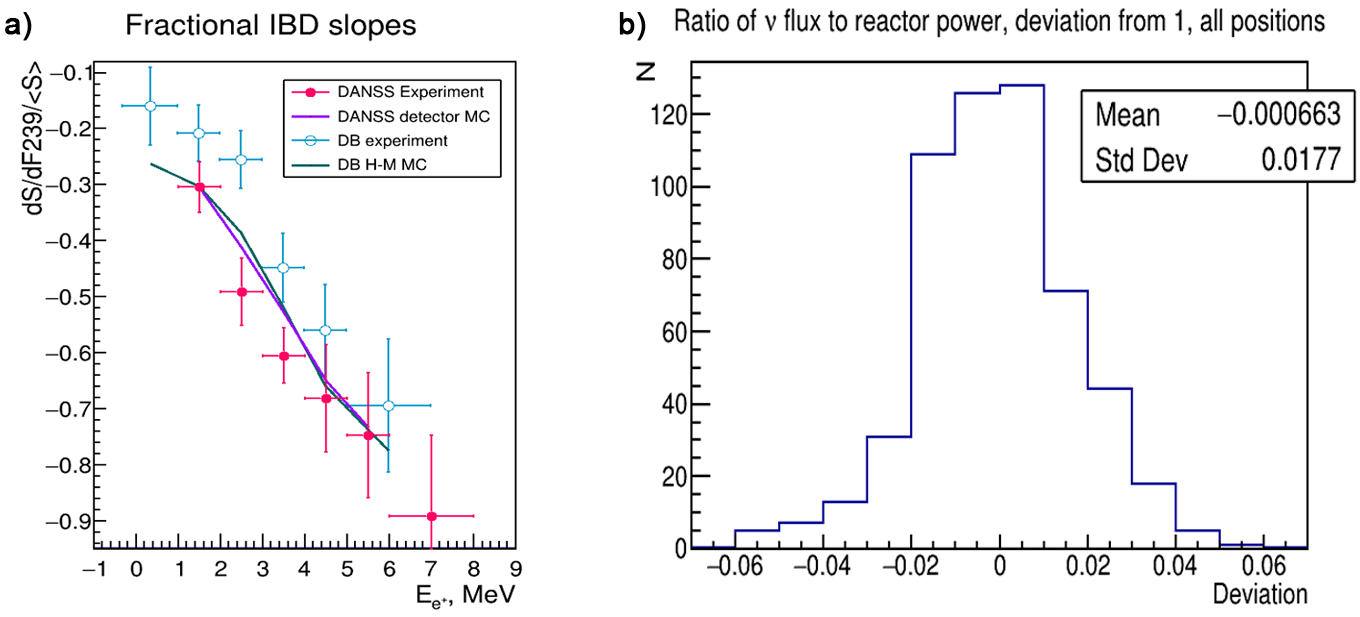}
\end{center}
\caption{\label{fig:power} 
a) Relative IBD rate dependence on the $^{239}$Pu fission fraction for different $e^+$ energies; 
b) Relative difference between IBD event rate in about 2 day periods and the reactor power during
the reactor operation at full power.} 
\end{figure}

 The relative IBD rate dependence on the $^{239}$Pu fission fraction for different positron energies (see Fig.~\ref{fig:power}a) agrees with the H-M model MC predictions and slightly larger in absolute values than the Daya Bay measurements
~\cite{An:2017osx}. 
The reactor power was measured using the anti-neutrino counting rate corrected for the changes in the efficiency and fuel composition according to the H-M model with the accuracy of 1.8\% in 2 days during almost 6 years
(see Fig.~\ref{fig:power}b). 
The IBD event rate was normalized to the reactor power during an initial one month period in 2016.
The rate at the different detector distances to the reactor core center was scaled taking into account the differences in the solid angle
that is almost proportional to  $1/L^2$.

\section{Search for sterile neutrinos.}

\begin{figure}[h]
\centering
\includegraphics[width=0.98\textwidth]{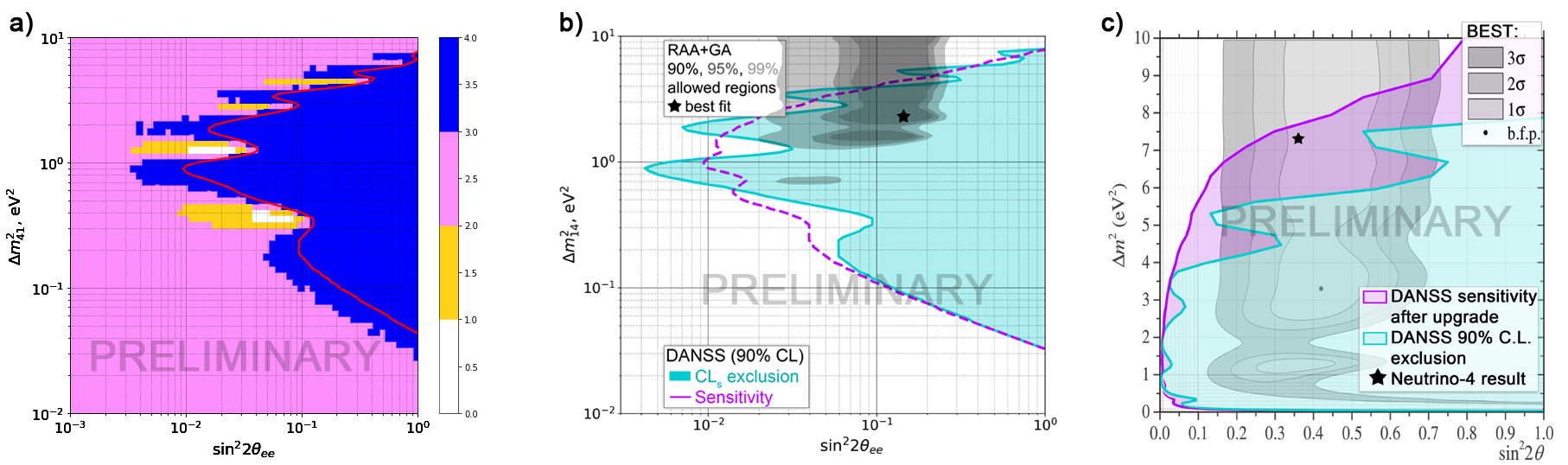}
\caption{a): Confidence levels for 4$\nu$ hypothesis at 1$\sigma$, 2$\sigma$, 3$\sigma$, and 4$\sigma$ obtained using the Feldman-Cousisns method, the red line corresponds to the 3$\sigma$ exclusion contour obtained using the Gaussian CL$_s$ method; b): 90\% C.L. exclusion area (cyan) obtained with the Gaussian CL$_s$ method in comparison with the RAA+GA expectations; c): The same exclusion area in comparison with the BEST results~\cite{BEST, BEST-2022-arxiv} and expected DANSS sensitivity after the upgrade.} 
\label{fig:exclusion}

\end{figure}

In the search for sterile neutrinos we use only ratios of positron spectra at the different distances from the reactor core. 
Systematic uncertainties in the energy scale and backgrounds were treated as nuisance parameters in the test statistics~\cite{RelEfficiency}. 
With additional 0.7 million IBD events collected in 2021 and 2022 the significance of the best-fit point in the 4$\nu$ case ($\mydm = 0.35$eV$^ 2$, $\mysin=0.07$) increased from  1.3$\sigma$ to 2.35$\sigma$. There is another point ($\mydm = 1.3$eV$^ 2$, $\mysin=0.02$) with a very similar significance.  The significance levels for the 4$\nu$ hypothesis obtained with the Feldman-Cousins method  are shown in Fig.~\ref{fig:exclusion}a. Still the indications in favor of the sterile neutrino are not statistically significant and we use the Gaussian CL$_s$ approach~\cite{CLS} to obtain  
the  exclusion area 
shown in Fig.~\ref{fig:exclusion}b that covers a very interesting range of the sterile neutrino parameters up to $\sin^22\theta_{ee} < 0.004$ in the most sensitive point. In particular a large part of the sterile neutrino parameters preferred by the recent BEST results~\cite{BEST,BEST-2022-arxiv} is excluded including the BEST best-fit point (see Fig.~\ref{fig:exclusion}c).

The DANSS collaboration has started the upgrade of the detector that should extend the sensitivity to higher $\mydm$ values and will allow to scrutinize even larger fraction of the BEST preferred area of the sterile neutrino parameters as well as  the Neutrino-4 result. However, the plans for the upgrade are delayed due to external problems.

The collaboration appreciates the permanent assistance of the KNPP administration
and Radiation Safety Department staff.
This work is supported by the Ministry of Science and Higher Education of the Russian Federation (Grant "Neutrino and astroparticle physics" No. 075-15-2020-778).


\begin{thebibliography}{99}
\bibitem{SAGE}
{Abdurashitov J. N. {\it et al.}}, \emph{{Measurement of the response of a Ga
  solar neutrino experiment to neutrinos from a $^{37}$Ar source}},
  \href{https://doi.org/10.1103/PhysRevC.73.045805}{\emph{Phys.Rev. C}
  {\bfseries {\bf 73}} (2006) 045805}.

\bibitem{GALEX}
{F. Kaether {\it et al.}}, \emph{Reanalysis of the gallex solar neutrino flux
  and source experiments},
  \href{https://doi.org/10.1016/j.physletb.2010.01.030}{\emph{Phys.Rev. B}
  {\bfseries {\bf 685}} (2010) 47}.

\bibitem{Huber}
{Huber P.}, \emph{Determination of antineutrino spectra from nuclear reactors},
  \href{https://doi.org/10.1103/PhysRevC.84.024617}{\emph{Phys.Rev. C}
  {\bfseries {\bf 84}} (2011) 024617}.

\bibitem{Mueller}
{Mueller T. A. {\it et al.}}, \emph{Improved predictions of reactor
  antineutrino spectra},
  \href{https://doi.org/10.1103/PhysRevC.83.054615}{\emph{Phys.Rev. C}
  {\bfseries {\bf 83}} (2011) 054615}.

\bibitem{Kopeikin:2021rnb}
V.I.~Kopeikin, Y.N.~Panin and A.A.~Sabelnikov, \emph{{Measurement of the Ratio
  of Cumulative Spectra of Beta Particles from ${}^{{235}}$U and ${}^{{239}}$Pu
  Fission Products for Solving Problems of Reactor-Antineutrino Physics}},
  \href{https://doi.org/10.1134/S1063778821010129}{\emph{Phys. Atom. Nucl.}
  {\bfseries 84} (2021) 1}.

\bibitem{Kopeikin:2021ugh}
V.~Kopeikin, M.~Skorokhvatov and O.~Titov, \emph{{Reevaluating reactor
  antineutrino spectra with new measurements of the ratio between $^{235}$U and
  $^{239}$Pu $\beta$ spectra}},
  \href{https://arxiv.org/abs/2103.01684}{{\ttfamily 2103.01684}}.

\bibitem{BEST}
V.V.~Barinov, B.T.~Cleveland, S.N.~Danshin, H.~Ejiri, S.R.~Elliott, D.~Frekers
  et~al., \emph{Results from the baksan experiment on sterile transitions
  (best)}, \href{https://doi.org/10.1103/PhysRevLett.128.232501}{\emph{Phys.
  Rev. Lett.} {\bfseries 128} (2022) 232501}.

\bibitem{BEST-2022-arxiv}
{Barinov V {\it et al. }}, \emph{A search for electron neutrino transitions to
  sterile states in the best experiment}, {\emph{arXiv:2201.07364v1} }.

\bibitem{Alekseev:2018efk}
{\scshape DANSS} collaboration, \emph{{Search for sterile neutrinos at the
  DANSS experiment}},
  \href{https://doi.org/10.1016/j.physletb.2018.10.038}{\emph{Phys. Lett. B}
  {\bfseries 787} (2018) 56}
  [\href{https://arxiv.org/abs/1804.04046}{{\ttfamily 1804.04046}}].

\bibitem{Danilov:2020ucs}
M.~Danilov, \emph{{New results from the DANSS experiment}},
  \href{https://doi.org/10.22323/1.390.0121}{\emph{PoS} {\bfseries ICHEP2020}
  (2021) 121}.

\bibitem{Ko:2016owz}
{\scshape NEOS} collaboration, \emph{{Sterile Neutrino Search at the NEOS
  Experiment}},
  \href{https://doi.org/10.1103/PhysRevLett.118.121802}{\emph{Phys. Rev. Lett.}
  {\bfseries 118} (2017) 121802}
  [\href{https://arxiv.org/abs/1610.05134}{{\ttfamily 1610.05134}}].

\bibitem{MiniBooNE2}
{Aguilar-Arevalo A A {\it et al.}}, \emph{{Significant Excess of ElectronLike
  Events in the MiniBooNE Short-Baseline Neutrino Experiment}},
  {\emph{Phys.Rev.Lett.} {\bfseries {\bf 121}} (2018) 221801}.

\bibitem{MicroBooNE:2021rmx}
{\scshape MicroBooNE} collaboration, \emph{{Search for an Excess of Electron
  Neutrino Interactions in MicroBooNE Using Multiple Final State Topologies}},
  \href{https://arxiv.org/abs/2110.14054}{{\ttfamily 2110.14054}}.

\bibitem{Serebrov:2018vdw}
{\scshape NEUTRINO-4} collaboration, \emph{{First Observation of the
  Oscillation Effect in the Neutrino-4 Experiment on the Search for the Sterile
  Neutrino}}, \href{https://doi.org/10.1134/S0021364019040040}{\emph{Pisma Zh.
  Eksp. Teor. Fiz.} {\bfseries 109} (2019) 209}.

\bibitem{Serebrov:2020kmd}
A.P.~Serebrov et~al., \emph{{Search for sterile neutrinos with the Neutrino-4
  experiment and measurement results}},
  \href{https://doi.org/10.1103/PhysRevD.104.032003}{\emph{Phys. Rev. D}
  {\bfseries 104} (2021) 032003}
  [\href{https://arxiv.org/abs/2005.05301}{{\ttfamily 2005.05301}}].

\bibitem{Danilov:2020rax}
M.V.~Danilov and N.A.~Skrobova, \emph{{Comment on \textquotedblleft{}Analysis
  of the Results of the Neutrino-4 Experiment on the Search for the Sterile
  Neutrino and Comparison with Results of Other Experiments\textquotedblright{}
  (JETP Letters 112, 199 (2020))}},
  \href{https://doi.org/10.1134/S0021364020190066}{\emph{JETP Lett.} {\bfseries
  112} (2020) 452}.

\bibitem{DANSS}
{Alekseev I. {\it et al.}}, \emph{{DANSS: Detector of the reactor AntiNeutrino
  based on Solid Scintillator}},
  \href{https://doi.org/10.1088/1748-0221/11/11/P11011}{\emph{JINST} {\bfseries
  {\bf 11}} (2016) P11011}.

\bibitem{DANSSdata}
{Alekseev I. {\it et al.}}, \emph{{Search for sterile neutrinos at the DANSS
  experiment}},
  \href{https://doi.org/10.1016/j.physletb.2018.10.038}{\emph{Phys.Lett.B}
  {\bfseries {\bf 787}} (2018) 56}.

\bibitem{Svirida:2020zpk}
{\scshape DANSS} collaboration, \emph{{DANSS experiment: current status and
  future plans}},
  \href{https://doi.org/10.1088/1742-6596/1690/1/012179}{\emph{J. Phys. Conf.
  Ser.} {\bfseries 1690} (2020) 012179}.

\bibitem{Bak:2018ydk}
{\scshape RENO} collaboration, \emph{{Measurement of Reactor Antineutrino
  Oscillation Amplitude and Frequency at RENO}},
  \href{https://doi.org/10.1103/PhysRevLett.121.201801}{\emph{Phys. Rev. Lett.}
  {\bfseries 121} (2018) 201801}
  [\href{https://arxiv.org/abs/1806.00248}{{\ttfamily 1806.00248}}].

\bibitem{An:2017osx}
{\scshape Daya Bay} collaboration, \emph{{Evolution of the Reactor Antineutrino
  Flux and Spectrum at Daya Bay}},
  \href{https://doi.org/10.1103/PhysRevLett.118.251801}{\emph{Phys. Rev. Lett.}
  {\bfseries 118} (2017) 251801}
  [\href{https://arxiv.org/abs/1704.01082}{{\ttfamily 1704.01082}}].

\bibitem{RelEfficiency}
{Skrobova N.}, \emph{Statistical data analysis in the danss experiment
  including antineutrino relative count rate data as a function of distance},
  \href{https://doi.org/10.3103/S1068335620090067}{\emph{Bull. Lebedev Phys.
  Inst.} {\bfseries {\bf 47}} (2020) 271}.

\bibitem{CLS}
{Qian X. {\it et al.}}, \emph{{The Gaussian CL$_s$ method for searches of new
  physics}},
  \href{https://doi.org/10.1016/j.nima.2016.04.089}{\emph{Nucl.Inst.Meth. A}
  {\bfseries {\bf 827}} (2016) 63}.
\end{thebibliography}

\end{document}